\documentclass[10pt,conference]{IEEEtran}
\usepackage[utf8]{inputenc}
\usepackage{cite}
\usepackage{amsmath,amssymb,amsfonts}
\usepackage{algorithmic}
\usepackage{graphicx}
\usepackage{textcomp}
\usepackage{xcolor}
\usepackage{xspace}
\usepackage{float}
\usepackage{balance}
\usepackage{ifthen}
\usepackage{listings}
\usepackage{url}

\usepackage{booktabs,mdframed,tabularx,multirow,mfirstuc}

\MFUnocap{are}
\MFUnocap{or}
\MFUnocap{a}
\MFUnocap{an}
\MFUnocap{and}
\MFUnocap{and/or}
\MFUnocap{as}
\MFUnocap{for}
\MFUnocap{but}
\MFUnocap{and}
\MFUnocap{the}
\MFUnocap{of}
\MFUnocap{on}
\MFUnocap{to}
\MFUnocap{at}
\MFUnocap{by}
\MFUnocap{to}
\MFUnocap{in}
\MFUnocap{if}
\MFUnocap{or}
\MFUnocap{with}

\newenvironment{smallquote}%
{\list{}{\leftmargin=0.15in\rightmargin=0.15in}\item[]}%
  {\endlist}
\newcommand{\quotebox}[2]{\begin{smallquote}\textit{``#2''}\ifthenelse{\equal{#1}{}}{}{ \mbox{-}~#1}\end{smallquote}}
\newcommand{\quoteboxtable}[2]{\textit{``#2''}\ifthenelse{\equal{#1}{}}{}{ \mbox{-}~#1}}

\newcolumntype{L}[1]{>{\raggedright\let\newline\\\arraybackslash\hspace{0pt}}p{#1}}

\newcommand{\code}[1]{\textsl{\capitalisewords{#1}}\normalfont}
\newcommand{\codeSC}[1]{\textsl{\capitalisewords{#1}}\sc}
\newcommand{\RQone}{What age- and gender-specific experiences have veteran software developers of marginalized genders had in their careers?}
\newcommand{\RQtwo}{What strategies have veteran software developers of marginalized genders adopted that they perceive as contributing to their survival in software engineering?}

\newboolean{showcomments}
\setboolean{showcomments}{true} 
\ifthenelse{\boolean{showcomments}}
  {\newcommand{\nb}[2]{
    \fcolorbox{gray}{yellow}{\bfseries\sffamily\scriptsize#1}
    {\sf\small$\blacktriangleright$\textit{#2}$\blacktriangleleft$}
   }
   
  }
  {\newcommand{\nb}[2]{}
   
  }

\newcommand{\todo}[1]{}
\renewcommand{\todo}[1]{{\color{red} TODO: {#1}}}

\newcommand{\change}[1]{#1} 
\newcommand{\changetoo}[1]{#1}

\begin{document}

\lstdefinestyle{custom}{
    belowcaptionskip=1\baselineskip,
    breaklines=true,
    frame=L,
    xleftmargin=\parindent,
    language=Java,
    showstringspaces=false,
    basicstyle=\footnotesize\ttfamily,
    keywordstyle=\bfseries\color{green!40!black},
    commentstyle=\itshape\color{purple!40!black},
    identifierstyle=\color{blue},
    stringstyle=\color{orange},
}

\lstset{escapechar=@,style=custom}

\title{``STILL AROUND'': Experiences and Survival Strategies of Veteran \change{Women} Software Developers}

\author{\IEEEauthorblockN{
Sterre van Breukelen\IEEEauthorrefmark{1}, Ann Barcomb\IEEEauthorrefmark{2}, Sebastian Baltes\IEEEauthorrefmark{3}, and Alexander Serebrenik\IEEEauthorrefmark{1}}
\IEEEauthorblockA{\IEEEauthorrefmark{1}Eindhoven University of Technology, The Netherlands, a.serebrenik@tue.nl}
\IEEEauthorblockA{\IEEEauthorrefmark{2}University of Calgary, Canada, ann.barcomb@ucalgary.ca}
\IEEEauthorblockA{\IEEEauthorrefmark{3}University of Adelaide, Australia, sebastian.baltes@adelaide.edu.au}
}
\maketitle

\begin{abstract}
The intersection of ageism and sexism can create a hostile environment for veteran software developers belonging to marginalized genders. In this study, we conducted \change{14} interviews to examine the experiences of people at this intersection\changetoo{, primarily women,} in order to discover the strategies they employed in order to successfully remain in the field. We identified 283 codes, which fell into three main categories: \textsl{Strategies}, \textsl{Experiences}, and \textsl{Perception}.
Several strategies we identified, such as \textsl{(Deliberately) Not Trying to Look Younger}, were not previously described \changetoo{in the software engineering literature}. We found that, in some companies, older women developers are recognized as having particular value, further strengthening the known benefits of diversity in the workforce. Based on the experiences and strategies, we suggest organizations employing software developers to consider the benefits of hiring veteran \change{women} software developers. \change{For example, companies can draw upon the life experiences of older women developers in order to better understand the needs of customers from a similar demographic.} 
While we recognize that many of the strategies employed by our study participants are a response to systemic issues, we still consider that, in the short-term, there is benefit in describing these strategies for developers who are experiencing such issues today.
\end{abstract}


\section{Introduction}
\label{sec:introduction}

\emph{``Congrats to all the young people but can we also give a massive shout-out to the older women in tech who have survived all the bullshit and are STILL AROUND\ldots''}\footnote{\url{https://twitter.com/triketora/status/1333899081656229888}}
This tweet alludes to the negative experiences and rarity of older women who are still active in the field of software.
While recent years have seen growing research attention to diversity in software engineering~\cite{Rodriguez-Perez21}, and both age~\cite{Kopec2018,Morrison,Baltes20} and gender~\cite{Terrell,CatolinoPTSF19} have been considered independently, the experiences of veteran women and non-binary people who have remained in the software industry have been rarely considered.

As the tweet indicates, the intersection of ageism and sexism can create a particularly harsh environment, and these experiences may not be adequately captured when examining sexism and ageism independently in the software industry. Software developers are overwhelmingly men. Women are estimated to account for approximately 25-30\% of employees the IT industry \cite{hill2010so}, but a recent survey of 70,000 developers conducted by Stack Overflow found that 93\% of developers were men \cite{StackOverflow2022}. By virtue of the well-established sexism in the field of software engineering (e.g., \cite{griffiths2010disappearing,smith2013working}), developers who are not men can be considered marginalized by gender.

While the same survey found that almost 33\% of developers were older than 35, the largest category was 25-34, accounting for almost 46\% of respondents \cite{StackOverflow2022}. Although the average age of the world-wide population of software developers is unknown, it is clear that older developers are marginalized \cite{Baltes20}, as exemplified by Mark Zuckerberg's famous 2007 quotation: \emph{``Young people are just smarter.''}

The challenges experienced by people at the intersection of multiple diversity aspects (e.g., age and gender) cannot be merely reduced to the combination of experiences associated with each one of the diversity aspects in isolation~\cite{Crenshaw1988}. For example, the experiences of Black women differ from those of Black men and of non-Black women~\cite{Ross2020}. In fact, Black women do not necessarily know whether their negative experiences should be attributed to their gender or their race~\cite{Thomas2018}.
\change{In this paper,} we were---in some instances---able to attribute particular experiences or strategies to `age' or `gender', but in many cases it was impossible to attribute observations to a single category, resulting in `age and gender' codes.

In our study, we looked at the experiences and survival strategies of people working in the software industry whose gender and age are marginalized, asking: 
\begin{itemize}
    \item RQ1. \RQone 
    \item RQ2. \RQtwo
\end{itemize}

We addressed these questions by conducting interviews with veteran software developers who identify as women or non-binary.
\change{Since we were only able to recruit one non-binary participant, and that person identified as women for most of their career, our results focus on veteran women developers, despite the the broader scope of our initial research questions.}
The majority of participants were still actively working as developers, although some had moved to managerial roles. 

Our contributions are twofold. 
First, we identify a number of strategies employed by \change{women} developers to succeed in the software industry.
These strategies range from specialisation and learning new technology to engaging in side projects and standing up for yourself, and from changing work environment and becoming consultants to changing appearance and unionizing.
Second, we find indications that older women developers have a particular value to the industry, which is recognized by some companies, e.g., because older women are an important part of the customer demographics.
Based on these contributions, we formulate several suggestions both for organisations employing veteran \change{women} software developers and for developers themselves.
While we acknowledge that many negative experiences that our participants reported are caused by systemic issues in the software industry, we still consider it valuable to report specific survival strategies of our participants, which might in short-term help developers facing similar issues.

\section{Related work}
\label{sec:rw}
Recent years have seen growing attention to diversity in software engineering~\cite{Rodriguez-Perez21,AlbusaysBDFMSS21,AdamsK20}.
\change{The homogeneity of tech culture in terms of age, ethnicity and gender has been widely recognized as a systemic issue
\cite{WHOAgeism21}.}
Most of \change{scholarship on software engineering} has considered  gender, often focusing on comparing women and men~\cite{Terrell,CatolinoPTSF19,Blincoe,VasilescuPRBSDF15,Lee}. 
More recently age, i.e., experiences of older adults, has been considered~\cite{Kopec2018,Morrison,schloegel2018age,DavidsonF}, with Baltes et al. questioning the notion of ``older'' in the context of software development~\cite{Baltes20}.
\change{
Morrison et al. studied veteran developers' role in open source and discussed differences to younger peers based on a panel discussion and quantitative study of Stack Overflow~\cite{MorrisonPMM16}, without including gender as a differentiating factor.
While Morrison et al. conclude with suggestions how veterans could contribute to open source, we studied the specific challenges and strategies of veteran women developers focusing on their experience in industry.
}

Several studies have combined gender and age by considering age when comparing gender-defined groups. Robles et al. 
have observed that women enter the free/libre/open source software (FLOSS) community later than men~\cite{RoblesRGD16}, while Wurzelova et al.~\cite{WurzelovaPB19} investigated age differences between women contributing to FLOSS and those not contributing to FLOSS.
In the domain of programming education, several studies have focused on experiences and perceptions of girls, implicitly combining gender and age~\cite{TELLHED2022107370,HappeBKW21,Aivaloglou2019}.

A number of authors have focused on older women entering  software development, e.g., Hyrynsalmi~\cite{Hyrynsalmi,Hyrynsalmi19} has studied women who switched to software development or planned on doing so. Seibel and Veilleux~\cite{seibel2019factors} compared factors influencing women entering software development through coding boot camps vs. computer science bachelor’s degrees, with the latter being older than the former.
As opposed to this line of work, our goal is to understand experiences of women that have been working in the software development industry for many years - and still are.

Attention to intersection between gender and age has been studied in the context of \change{the IT and communications (ICT) industry}.
Studies report discrimination based on both gender and age \cite{bandiasWarne}, in particular, in relation to motherhood~\cite{Griffiths06,griffiths2010disappearing}. 
On some occasions, older women have been made redundant~\cite{griffiths2010disappearing}, forcing them into self-employment.
Bandias and Sharma report that more women move to self-employment after the age of 54 and that it becomes more difficult for older women to achieve career advantages, despite having high technical abilities \cite{Bandias2016}. 
Beyond the world of ICT, studies report women employees either being seen as ``too young'' or ``too old'', mentioning age discrimination related to physical appearance and sexuality~\cite{duncan}. This belief was reflected in women feeling required to lose weight, wear high heels, and be more glamorous when they looked old \cite{handy2007gendered}. 
Opposed to this line of research, we also consider strategies veterans have used to overcome negative experiences and remain active in software engineering for an extended period of time. 
Our focus on software engineering rather than ICT or employment in general provides more focused insights.

\change{In 2021, the World Health Organization (WHO) published a Global Report on Ageism~\cite{WHOAgeism21}. In that report, the United Nations Secretary General writes that ``Ageism is widespread in institutions, laws and policies across the world. It damages individual health and dignity as well as economies and societies writ large.'' 
The report distinguishes institutional, interpersonal, and self-directed ageism. Our results are orthogonal to those levels of ageism, since we focus on the personal experiences of veteran software developers of marginalized genders. Such experiences can originate from all levels of ageism.
The report also acknowledges that being female or working in the tech industry increases the risk of being a target of ageism.
This underpins the importance of our study and the need for more research on ageism and sexism---and their intersection---in the software industry.}

Finally, attention to non-binary people has been limited.
\change{While there are a few studies focusing on non-binary developers~\cite{PradoMGP21, Gunawardena2022},} even studies that acknowledge presence of such genders~\cite{WangZhang2020,Crick2022} tend to exclude non-binary individuals from further analysis due to the limited amount of data available.
\change{While we did not exclude the non-binary participant we managed to recruit, they identified as woman for most of their career and thus their statements contribute to the challenges and strategies of women developers rather than to those specific to non-binary developers.}
\section{Methodology}
\label{sec:methodology}

To answer the research questions we opted for a series of exploratory interviews, also known as a qualitative survey~\cite{EmpiricalStandards}.
Since we could not build our study on a broad body of knowledge regarding gender and age in the software industry, we opted for an exploratory study design.
We chose semi-structured interviews because they allow eliciting unexpected information~\cite{HoveAnda,Seaman}.

As this is a human-subjects study, ethical approval was obtained from the Ethical Review Boards of our institutions, and
we follow the ethical guidelines of Strandberg~\cite{Strandberg19-0}.
After recruiting participants, we conducted interviews that were recorded, transcribed, and coded.
As a way of ensuring trustworthiness of our findings, we repeatedly compared emerging findings to the body of scientific knowledge and additionally performed member checking.

\subsection{Recruitment}
\label{ss:recruitment}

\begin{figure}[t]
    \centering
    \includegraphics[width=\columnwidth]{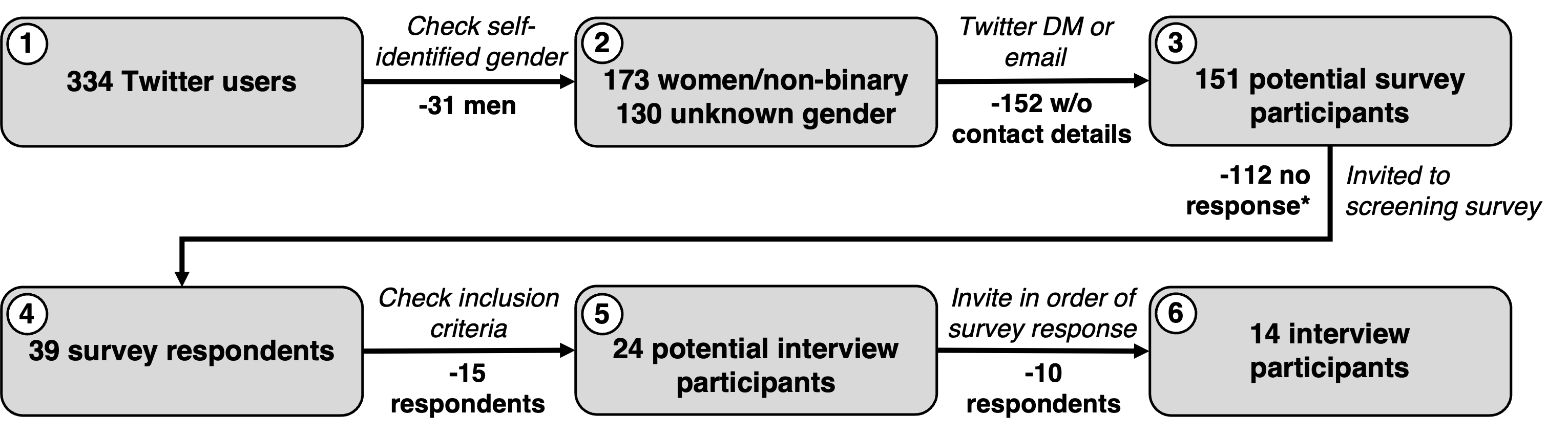} 
    \caption{Interview recruitment process (the asterisk indicates that the number might not be accurate due to snowball sampling via interview participants).}
    \label{fig:recruitment}
\end{figure}

Our target population was veteran software developers belonging to marginalized genders. This population is difficult to reach since there is no good sampling frame (i.e., there is no list of women and non-binary people in software engineering). We opted to select a `seed' group of participants by searching on Twitter for participation in threads \change{including} the one that sparked our investigation.
In the end, we identified enough participants from our original thread plus snowballing and thus did not include other threads.
We opted for Twitter over other social media such as TikTok, because Twitter has an older demographic \cite{ceci_2022,statista_research_department_2022} and provides APIs for researchers, which facilitated our search.
We started by identifying all Twitter users participating in the selected thread, operationalized as replying to the original message, replying to a reply, retweeting the original message, or replying to a retweet.
This yielded 332 Twitter users in our seed group (Step 1 in Figure~\ref{fig:recruitment}). 

We then looked at the user profiles, websites, and other tweets of participants in order to identify their self-identified gender, using their pronouns or use of gendered terms\footnote{For the entire list of terms, see \url{https://7esl.com/gender-of-nouns/}} (e.g., ``mother''). 
Among the 334 Twitter users, 31 appeared to identify as men, 173 as women or non-binary, and 130 provided no information from which we could infer gender. 
We excluded the users who identified as men (Step 2).

We contacted the remaining individuals using Twitter direct messages if available, or via email addresses provided in their user profiles or on websites linked from their profiles.
Our message directed respondents to a screening survey.
In that survey, we asked for gender, age, and number of years of experience in the software industry, in order to determine if the respondents were part of the target population.
We further asked if participants would be willing to participate in an interview, and, if so, for their email address.
This was necessary as our recruitment message did not contain a unique URL which would identify the person the invitation was sent to.
We were able to contact 151 Twitter users; for 152 users we could not retrieve contact information (Step 3).

Thirty-nine people responded to the survey (Step 4), of which 24 people met the inclusion criteria in terms of age, experience, and gender (Step 5).
We used 40 as the threshold of what constitutes an `older' developer based on the study of Baltes et al.~\cite{Baltes20}; the same threshold was used by Morrison et al.~\cite{MorrisonPMM16}.
As chronological age does not necessarily correspond to experience in the software industry, we focused on developers who had been in the industry long enough to accrue experiences and survival strategies.
Hence, we further operationalized `older' as having at least 18 years of experience in the software industry.\footnote{Based on a career starting at age 23, i.e., after schooling.}
To evaluate the stability of our findings, we also included three participants that, while not strictly belonging to the target demographics, might share experiences and strategies with other interviewees (see Table~\ref{tab:participants}).
We recruited one participant who was slightly younger than our target population, in order to determine if there was a \change{considerable} difference between our target population and slightly younger software developers. 
Similarly, we recruited one participant who left the industry, and one who left and subsequently returned, to evaluate whether our findings are likely to generalise to those individuals as well.
We further checked the gender of the survey participants, removing one participant who identified as man.
\change{We were unfortunately only able to recruit one non-binary participant. Since that participant identified as woman for most of their career, our results primarily apply to veteran women developers. We acknowledge that non-binary developers might face distinct challenges and develop different strategies, which future work needs to investigate in detail.}

We continuously checked the above-mentioned inclusion criteria for incoming survey responses.
The first author invited participants to an interview in the order of their survey response, until saturation was reached after interviewing 14 people (Step 6, see also Section~\ref{sec:qualitative-data-analysis}). 
We expected that members of the target population might know others within the target population, and thus we also employed snowball sampling by inviting interviewees to pass on our recruitment message \cite{FaugierSargeant,BaltesRalph}.

\subsection{Interview}

The interviews were conducted by the first author between November 2021 and January 2022. During an interview, we confirmed the participant's gender, age, and number of years working in the software development industry. We then asked about their experiences in the industry, especially those relating to gender and age, and about the behavioral strategies they adopted which they felt that younger or majority-gendered colleagues did not need to adopt. We began with open questions and then inquired about the specific strategies identified in the existing literature to determine whether they were recognized by participants.

The interviews were conducted online, using Microsoft Teams for recording and initial transcription. The transcriptions were corrected by the first author and were anonymized to exclude personally identifiable information such as names, company names, or exact locations. The names used to refer to participants in Table~\ref{tab:participants} and the remainder of this paper are pseudonyms. 
In some cases we have opted not to attribute a quotation in order to preserve the anonymity of participants.

\subsection{Qualitative Data Analysis}
\label{sec:qualitative-data-analysis}

We adhered to the coding guidelines of Saldaña~\cite{saldana_johnny_2016} and used the computer-assisted qualitative data analysis tool MAXQDA\footnote{\url{https://www.maxqda.com/}} to support the coding task.
We performed inductive coding~\cite{strauss1990open,thomas2003general}. 
Interleaved with the inductive coding, we performed axial coding~\cite{williams2019art}, grouping the codes in hierarchically organised themes.
Following Seaman~\cite{Seaman}, we recorded when modifications were made to the codebook, resulting in a list of codes gained from each interview. This allowed us to determine an a priori definition of saturation, which was influenced by the measures described by Guest, Bunce and Johnson~\cite{guest:2006:many}. We set an initial sample of six interviews, with a stopping criteria \changetoo{of two additional interviews}, and a 95\% threshold. 
If the codebook after six interviews already contained at least 95\% of the codes found after eight interviews, we determined that saturation had been reached. If the threshold was not reached, we coded \changetoo{one} more interview, then compared the state of the codebook after the \changetoo{ninth} interview against the state after the \changetoo{seventh} interview.
 In total, we conducted \change{14} interviews, although saturation was deemed to have been reached with the 10th and 11th interviews. 

\subsection{Data Availability}

We share the following artifacts as part of our supplementary material~\cite{supplementarymaterial}: ethics approval, recruitment messages, consent form, screening survey, interview script, final codebook exported from MaxQDA, and the scripts used for retrieving tweets, processing interview transcripts, and calculating saturation.
To not reveal the identity of our participants and to adhere to Twitter's guidelines as well as our ethics approval, we do not share the individual tweets, the list of Twitter users, the list of interview participants, and the interview transcripts.

\section{Results}
\label{sec:results}

\subsection{Demographics}

Fourteen people were interviewed between November 2021 and January 2022 (see Table~\ref{tab:participants}). 
\change{We used two pseudonyms for one participant due to the highly sensitive nature of the particular interview, as recommended by Strandberg \cite{Strandberg19-0}.}
In our sampling, we were primarily interested in age and experience, and our participants covered a broad spectrum within our target population. 
As explained in Section~\ref{ss:recruitment}, we also recruited three participants slightly outside that population.
These participants are presented in the lower part of  Table~\ref{tab:participants}.

\begin{table}
\centering
    \caption{Study Participants \change{(N$=$14; multiple pseudonyms were used to preserve the anonymity of one participant).}}
    \label{tab:participants}
    \begin{tabular}{l@{ }l@{ }l@{ }l@{ }l}
    \toprule

    \textbf{Pseudonym}& \textbf{Age}& \textbf{Experience}& \textbf{Gender}& \textbf{Status} \tabularnewline
    \midrule
    
    Alex & 55-59& 25-29& woman& active\tabularnewline
  Billie & 50-54& 18-24& woman& active\tabularnewline
   Bobbie&  60$+$& 25-29& woman& active\tabularnewline
    Charlie & 40-44& 18-24& woman& active\tabularnewline
    Dani & 60$+$& 45$+$& woman& active\tabularnewline
    Emery & 50-54& 18-24& woman& active\tabularnewline
     Erin &  60$+$& 45$+$& woman& active\tabularnewline 
    Jackie & 60$+$& 40-44& woman& active\tabularnewline
    Jaime & 50-54& 25-29& woman& active\tabularnewline
    Robin & 45-49& 18-24& woman& active\tabularnewline
    Sam & 40-44& 18-24& woman& active\tabularnewline
    Stevie & 60$+$& 40-44& woman& active\tabularnewline
    \midrule
    Elliot & 50-54& 18-24& woman& left\tabularnewline
    Noah & 40-44& 18-24& non-binary& returned\tabularnewline
    Riley & 35-39& $<$18& woman& active\tabularnewline  
    \bottomrule
    
    \end{tabular}

\end{table}

As our study concerns the intersectional effects of age and gender, we focused on these characteristics and did not explicitly look for other forms of diversity. However, some participants volunteered additional information during the interviews: one participant identified themselves as a person of color; one participant identified themselves as Korean; one participant identified themselves as transgender.
Because qualitative research seeks \change{to} describe experiences rather than quantify their frequency, the inclusion of participants with other experiences of marginalization helps ensure that the findings are more comprehensive. 

Several factors which we did not explicitly sample for were mentioned by participants: at the time of the interviews, 10 participants were located in North America; four participants identified themselves as parents; one participant mentioned having a PhD, two participants had a master's degree, and one participant did not finish their education. 
These factors might have affected the participants' workplace experiences. For example, women of color have to negotiate spaces of sexism and racism, often experience isolation, and lack social proximity to groups which are over-represented in technology leadership \cite{Thomas2018,twine2022geek}. 
Further, expectations concerning age vary by culture \cite{SchloegelSDM18} and mothers in male-dominated industries who are overworked are more likely to leave their occupations compared to employees without children \cite{cha2013}.

\subsection{Codebook}

The codebook consists of 283 codes, excluding those related to demographics.
As interviews of three participants not strictly belonging to our demographics did not results in different codes, we do not distinguish between codes derived from the interviews of these participants and from the interviews of the remaining participants.
At the top level, codes were categorized as \code{Experiences} and \code{Perception} (corresponding to RQ1), and \code{Strategies} (corresponding to RQ2). The following subsections discuss the results of each top-level code in more detail. \code{Strategies} was the largest category, with 135 codes. Under these categories, up to three sub-categories were used. An example of sub-categories is shown with the code \code{Energy wasted on fighting for a voice}, which occurs under \code{Perception} $>$ \code{Considered leaving} $>$ \code{Due to gender based experiences}.
A total of 1454 codings were applied across the 14 interviews.

\subsection{Experiences}

The top-level category \code{experiences} contributes to addressing our first research question: \textit{\RQone}

There were 72 codes relating to experiences, which are categorized into four main groups: \code{Age and/or Gender based experiences}, \code{Age based experiences}, \code{Gender based experiences}, and \code{General experiences}. Essentially, experiences which could be attributed to age or gender were applied to the relevant categories, whereas experiences which touched on both aspects were put in the combined category.
Experiences which did not appear to have gender or age dimensions were put in the general category.
In each of these experience sub-categories, codes were further divided into \code{positive} and \code{negative} experiences.

There were not many \code{positive} experiences related to age and gender, although \code{Being a role model} and \code{More opportunities due to gender and age} were found. One participant described how companies specifically looking to develop products aimed at her demographic led to opportunities:

\quotebox{}{A company approached me and said they were in the business, they wanted to make an app that would help predict who would have a stroke\ldots They were like `our ideal candidate would be a Woman of Color [who has] also survived a stroke.'\,}

\code{Negative} experiences were far more common, such as \code{Seen as non/}\code{less technical}, which has also been widely observed in the literature \cite{smith2013working,kenny}.

\quotebox{Sam}{I've been a director of game programming and at the senior software engineer [level] and I still get people who are trying to talk to me thinking that I won't understand technical concepts.}

Other participants highlighted how they experienced \code{Not being taken seriously} both when they were younger and when they were older. These examples suggest that women might face ageism at multiple stages during their careers, when they are `too young' and when they are `too old.'

\quotebox{Charlie}{I was discounted `cause I felt like people thought I was young.}
\smallskip
\quotebox{Emery}{As I approached menopause, there was another shift of just this contempt, because you're not even a sexually available female. And there's `No, I don't even have an interest in having sex with you and so why would I ever listen to you? You're going to try and tell me I'm wrong and you're unattractive.' So it got worse.}

\code{Age based experiences} included \code{positive} experiences such as \code{Wanted by company}, which is similar to \code{More opportunities due to age and gender}, but with an emphasis on social and intellectual capital acquired over a long career.

\quotebox{Noah}{I'm old enough to know that I can get another job. I'm old enough to know where I can put my skills. I have a network that I've built up.}

Some participants also experienced the opposite of \code{Not being taken seriously}, coded as \code{People listen due to age/}\code{experience}.

\quotebox{Robin}{I see people listening to me because I'm older.}

On the \code{negative} side, participants described being \code{Considered too old}, which led some to migrate toward management. Participants were frustrated and \code{Tired to keep proving yourself}.

\quotebox{Sam}{It's really exhausting to try to just sort of prove your existence over and over again. Or prove your competency over and over again. That's annoying.}

\code{Gender based experiences} included 26 codes. \code{Positive} codes included \code{getting opportunities} such as:

\quotebox{Bobbie}{I was always the first women, and people found that special and also encouraged me.}

Negative experiences included \code{toxicity in the industry}:
\quotebox{Dani}{Sometimes [there is] a subtle or not-so-subtle kind of hostile atmosphere, which can vary from simply not wanting to be in that kind of locker room environment to outright hostility.}
The largest group of negative experiences was related to sexism; it is summarized in Table~\ref{tab:sexism}.
\begin{table}
\centering

    \caption{Categories associated with the code \codeSC{Sexism}, following Leonard~\cite{leonard2021}.}
    \label{tab:sexism}
    {\renewcommand{\arraystretch}{1.2}
    \begin{tabular}{L{0.18\linewidth}p{0.73\linewidth}}
    \toprule

    \textbf{Code}& \textbf{Example}\tabularnewline
    \midrule
    \code{Ambivalent} & \quoteboxtable{Alex}{[there are] a lot of unwelcome words that are used towards women that actually exercised the same actions that men do.}\tabularnewline
    \code{Benevolent} 
    & \quoteboxtable{Stevie}{Especially when they start getting to `you're not technical enough.' I don't know how many men with the same equivalent skills as a woman in testing, get called not technical enough.}\tabularnewline
    \code{Hostile} & \quoteboxtable{Jaime}{I have been sexually harassed at most of my jobs up until I turned 40.}\tabularnewline
    \code{Institutional} 
    & \quoteboxtable{Jackie}{I've had some clients who said if you were a male consultant, I would certainly pay you what you're asking, but you're female.}\tabularnewline
    \code{Internalized} 
    & \quoteboxtable{Billie}{I currently coach another woman on my team. She is a textbook  discriminated woman, but she attracts it.}\tabularnewline
    \code{Interpersonal} & \quoteboxtable{Elliot}{People would tell you to your face that like you needed to be sweeter. You know, like you needed to be more feminine and you \change{needed} to be pretty or you know all of those things.}\tabularnewline
    \bottomrule
    
    \end{tabular}}

\end{table}

Under \code{general experiences} there were \code{positive} codes associated with \code{related to the type of work}. Multiple participants talked about problem-solving as a key reason for enjoying the job. \code{working with smart/}\code{nice} was another common code in this category.

\quotebox{Jackie}{I really enjoyed solving problems in a way that people would use their product.}

Among the \code{negative} experiences were other forms of discrimination and hostility, both observed and experienced.

\quotebox{Alex}{I am at this point one of two women left standing in tech, and this kind of tells me that the industry is still very toxic for women and People of Color.} 
\smallskip
\quotebox{Charlie}{They had all these diversity training and workshops around identity and racial equity. And I have never witnessed [a] more hostile workplace towards transgender people, Black people, and women.}

There were also other \code{toxic/}\code{difficult work environment} experiences, some involving \code{high work load/}\code{overtime}:

\quotebox{Sam}{I've definitely encountered some really toxic work environments. I've had a lot of overtime. \ldots So stress, exhaustion, there's been some really brutal working environments.}

Finally, there were experiences around \code{pregnancy/}\code{parenthood}, some of which were \code{positive}:

\quotebox{Alex}{The positive interactions I've had were with managers that understood that my family was important to me.}

Most of the experiences, however, were \code{negative}. These included \code{discrimination/}\code{harassment}, \code{excluded from after work activities due to parenthood}, \code{lack of understanding from managers}, and \code{lack of work-}\code{life balance}.

\quotebox{Jackie}{The very first time I was fired from a director level position. I was 38 weeks pregnant, like 2 weeks away from my due date and my bosses hired a new VP and he came in and took, literally, took one look at me and said `they didn't tell me you were pregnant.' I was a director \change{of} software development. \ldots They said, `well, no, we're gonna start talking about a package' which meant that was a severance package, a lay-off which meant I knew I was going to make a whole lot of money. `Cause he would not have fired a man for being pregnant or having a pregnant spouse.}

\subsection{Perception}

Together with \code{experiences}, \code{perception} completes addressing our first research question by extending the more objective experiences with the subjective experiences (perceptions) participants had.
The \code{Perception} category included the following sub-categories: \code{feeling towards leaving}, \code{value}, \code{current work place}, \code{industry}, and \code{task assignment}.

The category \code{feeling towards leaving} covered both participants who \code{considered leaving} and those who had \code{not considered leaving} the industry. In eight interviews the theme \code{considered leaving} was present, while in 10 interviews the theme \code{not considered leaving} occurred. This is because some participants expressed more complex feelings, such as not feeling like leaving at present, but having had the feeling in the past. Table~\ref{tab:considered-leaving} shows the sub-codes immediately under the codes \code{considered leaving} and \code{not considered leaving}.

\begin{table}
\centering
    \caption{Sub-codes associated with the code \code{considered leaving}.}
    \label{tab:considered-leaving}
    {\renewcommand{\arraystretch}{1.2}\begin{tabular}{L{0.25\linewidth}p{0.65\linewidth}}
    \toprule

    \textbf{Code}& \textbf{Example}\tabularnewline
    \midrule
    \multicolumn{2}{l}{\code{considered leaving}}\tabularnewline
    \midrule
    \code{age based experiences} & 
    \quoteboxtable{Alex}{The older I get, the less I wanna deal with. The young girl I was, it was like I had all this energy. I can do all this stuff. Yeah, sure, I’ll work on more hours than I should. Yeah, I’ll put all this stuff on hold. Yes, I’ll do this. I won’t do that anymore. \ldots Life is too short.}
    \tabularnewline
    
    \code{gender based experiences} & \quoteboxtable{Emery}{When you are just spending all your energy into fighting people trying to knock you down and people who just want you to not have a voice and to not make decisions. \ldots It’s so exhausting.}\tabularnewline
    
    \code{industry based experiences} & \quoteboxtable{Alex}{Many times whenever I run into a block in my career when there are megalomaniac managers. When they were, whenever there are leaders of a company that just aren't inspiring to me.}\tabularnewline
\midrule
    \multicolumn{2}{l}{\code{not considered leaving}}\tabularnewline
    \midrule
    \code{enjoys the work} & 
      \quoteboxtable{Elliot}{No. I love this business.}
    \tabularnewline
    \code{miss the people} & \quoteboxtable{Robin}{But I would miss all the geeks.} \tabularnewline
    
    \code{miss the positive atmosphere} & \quoteboxtable{Robin}{My sister goes, how can you have so many stickers on your computer when you just started in this job? And that you know, we have geeky T-shirts that we discuss anything from.} \tabularnewline
    
    \code{more in control} & \quoteboxtable{Noah}{Now I just feel so much more in control of my career, and so I think about leaving. But I think about it in a sense of like this sabbatical, you know, maybe I'll take a year and I will fix my house and get some sleep.}
    \tabularnewline
    
    \code{pays well} &
    \quoteboxtable{Jaime}{I'm in this job because  when I was young I looked at what was open to me with my various aptitudes and  I am going to have a happy life with enough money.}
 \tabularnewline
    
   \code{to be a good role model} & \quoteboxtable{Charlie}{To have women in this industry who are in their 50s and 60s, and they’re writing code \change{and} shipping code and keeping their skills like modern and like that’s been huge to me and I want to be that for other people} \tabularnewline

    \bottomrule
    
    \end{tabular}}

\end{table}

In the category of \code{value}, there were both positive codes, such as \code{well valued}, and negative codes, such as \code{undervalued}. Both the positive and the negative aspects of value could be applied to multiple evaluators, such as \code{by boss}, \code{by clients}, and \code{by peers}. The importance of work being recognized in order to progress in a career is illustrated with a positive and a negative example.

\quotebox{Charlie}{I just got a call today from a woman who is like, ‘I loved what you were doing. Would you consider coming to work at my startup?’ and it feels good to get new opportunities coming towards you. So I feel like my work that even wasn't successful was valued.}
\smallskip
\quotebox{Charlie}{I said there is no other project manager on this floor in this department who has got the same financial like wins for the bottom line, and if I were a man, I would be given this role that is much higher. But no, I’m not getting picked for those things. \ldots I was asked for ten years of my career history in that justification.}

One observation about the \code{current work place} was that participants appeared to prioritize good working environments as they gained experience. 

\quotebox{Robin}{So we actually have a cultural interview as well as a competent interview, and that has the same value. So I chose. I've chosen a company where of course, we still make mistakes, but where we believe in being human.}

When speaking about the \code{industry}, participants also highlighted the \code{importance of a good environment/}\code{atmosphere}. 

\quotebox{Jackie}{Oh yeah. I mean. I get to create the atmosphere now with my clients and in my classes and all that stuff and that really helps. But yeah, certainly the environment is a huge huge thing.}

Respondents had mixed views about \code{age and gender in the industry}, \code{aging in the industry}, and \code{gender in the industry}.

\quotebox{Billie}{As women, it is still the fact that we have to be louder and \change{to} achieve more to be acknowledged.}

\quotebox{Bobbie}{I think the gender is not really  a \change{disadvantage}, maybe it's an advantage because people like you better. If you were a female, like with an opinion, that people really accept more.}

\quotebox{Elliot}{I mean, the age thing I would say has been more of a positive than a negative.}

\quotebox{Dani}{One of the few acceptable ways to be a woman in computer science is to be the local office decoration. And I'm not very decorative at 62.}

We identified 18 mentions of \code{task assignment} involving \code{assigned differently}, which could be due to age or gender, and six mentions of \code{task assignment} being the \code{same as others}. Examples of different task assignment could relate to being relegated to menial chores typically assigned to women:

\quotebox{Alex}{I feel like I’m usually assigned a task to get everyone together. Take notes. Organize a celebration. Uhm, you know all of the social stuff that never gets you ahead.}

However, the tasks could be more appealing, as the participant became better known in the organization---but still not necessarily beneficial for the individual's career:

\quotebox{Noah}{As I’ve gotten more experienced, I get the opposite [less glamorous], which is that people hand me things that they don’t know if this is a good idea. They don’t know if it’ll work, they don’t actually know if anybody can do it, and I love those kinds of challenges, but I also know that they’re kind of pitfalls, right? Because you know you put somebody in that position. They’re a little bit expendable. You’re picking a person that you’re OK with them failing.}

\subsection{Strategies}

The top-level category \code{strategies} directly addresses our RQ2: \textit{\RQtwo}

\code{Strategies} contained the following second-level categories: \code{age related strategies}, \code{gender related strategies}, \code{age and gender related strategies}, and \code{general strategies}. These categories were used in much the same way as the categories of the same name under the top-level code \code{experiences}: the most specific category which could be applied was used.

Under \code{age and gender related strategies} there were four categories. \code{against the bias} focuses on strategies of going against the gender and age bias experienced by participants. It contains sub-codes such as \code{standing up for yourself} and \code{gathering allies}. The code \code{side projects} involves relevant activities outside of employment, such as \code{serving on board}, \code{informing others/}\code{mentoring/}\code{coaching}, and \code{speak (at conferences)}. Another approach is \code{changing work environment}, which could be expressed in several ways: \code{changing companies} $>$ moving to a \code{different company}, \code{changing companies} $>$ starting your own \code{new company}, \code{changing roles} $>$ \code{consulting}, and \code{changing roles} $>$ \code{moving up the chain} of command. Finally, \code{career related} had the sub-codes \code{networking}, \code{take sabbatical}, \code{using your influence}, and \code{using/}\code{leveraging experience}. Table~\ref{tab:strategies-age-gender} depicts an example strategy from each of the four categories under \code{age and gender related strategies}.

\begin{table}
\centering

    \caption{Categories associated with the code \code{age and/or gender strategies}.}
    \label{tab:strategies-age-gender}
    {\renewcommand{\arraystretch}{1.2}\begin{tabular}{L{0.25\linewidth}p{0.65\linewidth}}
    \toprule

    \textbf{Code}& \textbf{Example}\tabularnewline
    \midrule
    \code{against bias} & \quoteboxtable{Dani}{All those young guys don’t want their mom programming with them, their grandmother, on the other hand.}\tabularnewline
    
    \code{side projects} & \quoteboxtable{Elliot}{Try to do what we can you know for the younger women and younger non-binary people and you know the younger trans people.}\tabularnewline

    \code{changing work environment} & \quoteboxtable{Robin}{I’m lucky because I don’t have [a] mortgage. I don’t have anything I need to pay. If everything goes really terrible, I can stay with my friends for a while and get another job. So that way I’m lucky because if you have kids, family, you’re in [a] different position.}\tabularnewline 
       
     \code{career related} & \quoteboxtable{Noah}{And I mean gender does play into it, in that you know a lot of my network started. With people of marginalized genders supporting each other.}\tabularnewline
  
    \bottomrule
    
    \end{tabular}}

\end{table}

In \code{age related strategies} we also identified a number of different categories. \code{career related strategies} included codes such as \code{find a niche}, \code{having a back up plan} and \code{set boundaries}. The following quotation illustrates \code{find a niche}, which suggests that specific industries within software development might be more open to older developers:

\quotebox{Dani}{I have become very specialized. Fortunately, I have found a niche, [in] which most programmers are older.}

The most common sub-code was \code{teaching yourself (}\code{new) tech}, with 11 coded segments describing how participants continuously taught themselves new technology even when their employer did not see the value in it.

\quotebox{Alex}{But they said, oh, I wasn't technical enough to do this or that, so I went on the web and just like, if they’re telling me this, let me brush up on system design, and so I did that.}

Another \code{age related strategy} involves \code{behaving younger} by either \code{changing the way you speak/}\code{vocabulary} or \code{fit in with company culture}.

\quotebox{Sam}{I do tend to pick up. You know the colloquialisms, the slang, whatever. I learned to speak younger somewhat deliberately.}

In addition to \code{behaving younger}, some participants adopted the strategy of \code{changing appearance}, which could include, among other things, actions such as \code{changing hair to look younger}, or taking an opposite tack of \code{(}\code{Deliberately) not trying to look younger}.

\quotebox{Jaime}{I don’t do this, but I know I have friends who have gotten Botox in order to look younger. Um, I think it’s fairly common.}

Participants also employed some of the strategies used for \code{age and gender related strategies}, such as \code{changing roles}.

\quotebox{Alex}{When I was midway through my career, I was seeing younger people coming in, which is fine. I wasn’t that old, but \ldots I would have to learn all this stuff all over again. And so I got to a point where I don’t wanna do this anymore. \ldots  I said, `If I stay as an individual contributor, I think I only have more problems down the line getting the salary and the positions that I want.’\ }

We found that \code{gender related strategies} contained the most strategies, with eight separate categories and 308 code segments. The categories were: \code{against gender bias strategies}, \code{career related strategies}, \code{changing work environment}, \code{changing your appearance}, \code{communication methods}, \code{ignoring situations}, \code{traditionally feminine}, and \code{traditionally masculine}. Of these, \code{against gender bias strategies} was the largest category, with 70 code segments and eight sub-codes, such as \code{backing other women up}.
Table~\ref{tab:strategies-gender} provides an example of a strategy from each of the categories.

In the category of \code{general strategies}, we found strategies which are not necessarily associated with age or gender, such as \code{unionize} and \code{knowing your worth}. Because these strategies do not explicitly relate to the research questions, a single example is used to illustrate this category:

\quotebox{Jaime}{I also see a lot of movement towards unions, which I like. I think that's a good thing. I think we need to force that. \ldots I think it's healthy, and I think that would it would do a lot toward rationalizing salaries.}

As earlier studies reported software developers doubting appropriateness of unions to their work and their labor situation~\cite{Barett2001,BaldryEtAl2005}, future studies should investigate  whether the positive attitude of Jaime (and a similar attitude of Robin) are common among older developers of marginalized genders.

\begin{table}
\centering

    \caption{Categories associated with the code \code{gender related strategies}.}
    \label{tab:strategies-gender}
    {\renewcommand{\arraystretch}{1.2}\begin{tabular}{L{0.25\linewidth}p{0.65\linewidth}}
    \toprule

    \textbf{Code}& \textbf{Example}\tabularnewline
    \midrule
    \code{against gender bias strategies} & \quoteboxtable{Elliot}{You need to stand up for yourself because men are different. Generally, in my experience, if you don’t stand up to men, they're gonna squash you. If you stand up to them, even if they think you're [a] bitch. They're gonna respect you more.}\tabularnewline
    \code{career related strategies} & \quoteboxtable{Emery}{I went to a women in tech talk once and there were younger women on the panel, and they were in sales, and they were saying ‘Oh yeah, my colleagues. They steal my idea, \ldots but it’s OK because at the end of the day we all split the commission’. And I’m like, ‘no, that's not true. You are wrong!’ Because, what has happened is that the end of the day, he's getting the promotion for stealing your idea, or he's getting the promotion because it seems like he has the better client relationship.}\tabularnewline
    \code{changing work environment} & \quoteboxtable{Sam}{We're trying to get opportunities for people to work. So we've hired more female presenting engineers and some of our goals to get me to actually be there lead on projects because it's extremely rare to actually have.}\tabularnewline
    \code{changing your appearance} & \quoteboxtable{Robin}{I know I have female friends, women friends who will never speak at a conference in a short dress or showing your cleavage. Because they will get harassed because they look good.}\tabularnewline
    \code{communication methods} & \quoteboxtable{Jaime}{Yeah, pretending that I'm not direct. (laugh) Seriously, I softened my language so much when I was a kid. I was a consultant. \ldots I'm very polite and I'm very diplomatic. I've retained that from my consulting years, but it's - I think I read a room very, very well. Because and that's that's what I've noticed. From a lot of women we read a room very well because, it's necessary.}\tabularnewline
    \code{ignoring situations} & \quoteboxtable{Stevie}{Yeah, I do my share of ignoring. Lalala, I don't wanna know about it.}\tabularnewline
    \code{traditionally feminine} & \quoteboxtable{Emery}{I mean when you are passive, you don't get made a target. That is definitely the reason that a lot of people choose that path.}\tabularnewline
    \code{traditionally masculine} & \quoteboxtable{Riley}{I used to present myself in a very traditionally male way. And you know, I adopted a number of traditionally male hobbies and things that I enjoyed in a way of speaking, um to, to mask into to help myself fit in better.}\tabularnewline
    \bottomrule
    \end{tabular}}
\end{table}

\section{Discussion}
\label{sec:discussion}

Our research found that veteran \change{women} software developers had many negative experiences linked to age, gender, or both age and gender, as well as some positive experiences attributable to age and/or gender. They also observed and developed many strategies which helped them persist in the industry despite high attrition rates. In this section, we look at some of the more prominent experiences and perceptions, and the strategies which relate to them, in the context of existing literature.

\subsection{Visibility}

Many of the strategies and experiences mentioned throughout the interviews seem to connect to visibility or the lack thereof. Visibility is critical for career progression, and a lack of visibility might lead women to lose out on promotions \cite{correll2016}. Another reason for lack of promotion might be that supervisors are unlikely to attribute women's performance to skills rather than luck \cite{lemons2001}. This could be why some participants in the interviews emphasize getting credit for their work. One of the participants directly connected men stealing ideas with women being overlooked. Additionally, noting down accomplishments might allow people to recognize their own capabilities, and make supervisors aware of them. 
Some suggest that visibility can be increased by getting involved with challenging assignments \cite{ragins}.
However, some participants found that they were assigned career-limiting tasks. The strategy of \code{volunteering for tasks} was a way of ensuring more challenging and meaningful work, another was \code{setting boundaries}, both of which were \code{career related strategies} under \code{gender related strategies}. 

Other ways of increasing visibility are social events such as conferences, and networks \cite{vinnicomb}. We also see these and similar strategies:  \code{speaking at conferences},  \code{networking} and \code{gathering allies}. \code{side projects} can be another way of networking and increasing visibility; one participant directly attributed these strategies to career progression. Networking can help  overcome the feeling of exclusion, and can provide a sense of identity. Networks have often been used to exclude women \cite{griffiths_moore}, who are outside the `old boys network'. In response, all-women networks have been established \cite{berkelaar}. Participants mentioned both formal and informal networks for under-represented genders as a way of combating exclusion. Because of strong networks, some participants felt confident in \code{changing work environment}.

Marginalized genders can also be too visible, in the sense that they are expected to conform to gendered behavioral norms. Tyler and Cohen  reported a participant as saying, ``I think as a woman you’re expected to be always happy, happy, shiny, accessible for people to come and talk to'' \cite{tyler2010spaces}. We observed a similar pressure for women to present themselves according to feminine stereotypes:

\quotebox{Charlie}{[The CEO] was like, `Are you OK?' And I was like, `I am definitely OK, why?' And he's like, `Are you sure? I mean, you didn't smile one time in that meeting.'\ }

\subsection{Appearance}

We saw \code{Changing appearance} concerning being traditionally feminine. More than half of the mentions regarding \code{less feminine/}\code{more gender neutral} happened in early career. The reason that some women feel like they need to dress less femininely could be related to stereotypes. Sczesny et al.  performed two experiments to determine how physical appearance and sex influence gender stereotypes \cite{Sczesny}. They found that physical appearance influences gender stereotypes concerning leadership competence. Thus, those who presented with a masculine appearance were ascribed higher leadership qualities than those with a feminine appearance, regardless of their sex  \cite{Sczesny}. Fleischmann et al.  support the findings that feminine appearance activates female gender stereotypes \cite{fleischmann2016}. Additionally, they found that participants assumed women in feminine outfits would be less successful in solving computer tasks than women in gender-neutral clothing. Moreover, they found that participants also estimated that women dressed femininely would take longer to complete the task and rated the computer skills of the women in feminine clothing lower than those in gender-neutral outfits \cite{fleischmann2016}. These findings might explain why some women software developers actively change their dress to be more masculine or gender-neutral. Firstly, they may adopt styles to be perceived as more competent and seen as more leadership material. This would also align with the findings that several women decide to dress more femininely in their later careers as several women have moved to leadership/management positions and might not feel the need to conform anymore. Secondly, in their later careers, women might feel more confident in their skills and care less about the assumptions of others.

While women may experience pressure to appear more traditionally masculine early in their careers in order to be taken seriously, as they age, they  are pressured to appear younger and physically attractive. Several participants mentioned dying hair and Botox as a way of conforming to expectations.

\quotebox{Jaime}{I also have a friend who gets Botox so that she doesn't look angry on video calls.}

Others, however, responded by \code{(}\code{deliberately) not trying to look younger}, which, combined with \code{against bias} suggests that there may be a point at which pressures related to appearance diminish.

\quotebox{Dani}{I have found more luck now that I'm old enough to do it with with actually like. Maybe you know with sort of being the old lady? Because that gets, in some sense, it gets me out of the whole thing.}

\subsection{Leaving}

We have discussed several different aspects that play a role in why people feel like leaving. The literature discusses why women leave~\cite{WilliamsDempsey}, and industry plays a large role. The ICT sector is known for its overtime, heavy workloads, and pressure of deadlines, which can be a source of frustration and exhaustion \cite{hanappiegger2012}. Women also face more sexism, workplace bullying, a pay gap, a glass ceiling, and difficulty being accepted as supervisors \cite{hanappiegger2012}. Studies have indicated that women must work harder than men because women are held to stricter working standards~\cite{Gorman}. Additionally, there is still an expectation that women will perform more housework, which might conflict with work expectations \cite{hanappiegger2012}. Our participants mentioned work-life balance, not being taken seriously, and feeling excluded as reasons for considering leaving. When women leave their profession, they tend to leave the industry, thus suggesting they view the issues as being industry-wide, not specific to a company \cite{hanappiegger2012}. One participant questioned the constant need for women to prove themselves in the software industry:

\quotebox{Emery}{Why do all women have to be a superstar? What's wrong with being mediocre?}

The two participants who left software development left the industry completely, but also returned at some point - suggesting that there was a method to return and that leaving does not have to be forever. Also, while several participants considered leaving, many of them ultimately chose not to, for reasons which included \code{enjoys the work}, \code{miss the people}, and \code{to be a good role model}, among others.

\quotebox{Alex}{The thing that kept me in so long was my daughter. I didn’t want to quit because I didn’t want her to quit.}

This suggests that the strategies we identified may be able to help \change{women} software developers navigate the industry as it is today. However, it is important to recognize that the issues of ageism and sexism are systemic, and companies and the industry as whole need change to retain the experience and perspective of veteran developers. When people leave an industry, knowledge is lost, regardless of gender. However, when an industry loses many people of a specific group, their particular knowledge is lost. Firstly, women's lived experience is different from men's; thus, their perspective \change{brings} unique views \cite{crump2000women}. Additionally, research shows that gender diversity in R\&D teams leads to radical innovations \cite{diazGarcia}, and different points of view \cite{QUINTANAGARCIA2008492}. The loss of knowledge is not the only negative; the loss of diversity is another negative aspect. Several studies have observed positive aspects of diversity in software development teams \cite{VasilescuPRBSDF15, CatolinoPTSF19, diazGarcia, Blincoe, OSTERGAARD2011500}. Some of our participants also found that, because of their relative rarity, they were especially valued:

\quotebox{Elliot}{Now, it’s interesting because the gender and age has combined because the actual businesses are more oriented towards older women. And, I did not expect that. But because there’s so few older \change{women}.}

This finding suggests that companies who do not currently have an age- and gender-diverse pool of developer talent are missing opportunities to better understand  the needs of customers and society as a whole. 

\subsection{\change{Gender Biases}}
\change{In their well-known book ``What Works for Women at Work,'' Williams and Dempsey have identified four gender biases, i.e., patterns of behavior forming major obstacles to women's advancement to leadership~\cite{WilliamsDempsey}: `Prove It Again', `Tightrope', `Maternal Wall', and `Tug of War'. 
Imtiaz et al.~\cite{DBLP:conf/icse/ImtiazMCRBM19} operationalized these biases and evaluated their presence on GitHub.
Given the omnipresence of the biases claimed by Williams and Dempsey, it is not surprising that some of the experiences reported by our interviewees can be associated with them.}

\change{`Prove It Again,' the recurrent demand for women to prove their ability to be considered competent, is related to the interviewees reporting it was \code{tiring to keep proving yourself}. 
The `Tightrope' is related to the narrow space women have between adopting more traditionally feminine ways and exacerbating the `Prove It Again' problems, and adopting more traditionally masculine ways and being seen as lacking social skills. 
We observe a similar phenomenon with women adopting \code{traditionally feminine} and \code{traditionally masculine} strategies. 
Gender-related strategies include adjusting their appearance (\code{changing your appearance}, with sub-codes such as \code{dressing strategically}). 
Interviewees also adopted age-related strategies pertaining to \code{change appearance} which may have a gendered component, such as \code{plastic surgery/}\code{botox}. 
However, we also observed participants deliberately rejecting pressure regarding their appearance, with the gender-related strategy \code{to be memorable} and the age-related strategy \code{(}\code{deliberately) not trying to look younger}. 
The `Maternal Wall' puts mothers in double bind of either spending more time at work and being seen as a bad mother or spending less time at work and being pushed to the margins of their profession. 
Our interviewees discuss \code{pregnancy/}\code{parenthood related experiences}, 
and while some of them positive (\code{flexibility from job}), 
 most of them are 
 negative (e.g., \code{lack of work-}\code{life balance}). 
The `Tug of War' leads ``women to judge each other on what's the right way to be a woman'' often causing more senior women to treat more junior women more harshly than men.
This bias did not emerge during our interviews; rather we saw more positive interactions such as \code{informing others/}\code{mentoring/}\code{coaching}, 
\code{creating safe spaces}, 
and \code{gathering allies}. 
}

\subsection{Actionable Insights}
The actionable insights of this work are twofold. First, we would like to raise awareness of managers for the issues older women and non-binary individuals in their companies are facing.
Second, based on our study, we formulate several suggestions both for organisations employing the developers and for developers themselves.

\subsubsection{Organisations}
Organisations should invest in creating a good working environment and a positive atmosphere, investing older developers of marginalized genders with sense of control of their work and their careers, supporting their promotion, assigning tasks, and paying them on par with men.
While these recommendations are true for any employer, they are even more pertinent for software engineering given the scarcity of older women and non-binary people in this industry.
Moreover, specifically in case of software engineering, the inclusion of developers who are more representative of the population ensures that the software can meet the needs of society as a whole.
Gender diversity is beneficial for productivity of software development teams~\cite{VasilescuPRBSDF15} and sense of control for older workers is associated with maintaining well-being~\cite{Liu2017}, which is one of the aspects of developer productivity~\cite{ForsgrenSMZHB21}.

\subsubsection{Developers}
Developers themselves can move to a different work environment (e.g., by starting their company, moving to a different company, becoming consultant or manager) or try to change their work environment (e.g., by unionizing, standing up against gender bias, or carving new opportunities for themselves).
Changing appearance is one of the commonly mentioned but profoundly problematic strategies.
We recognise that these recommendations are merely band-aid solutions applied to systemic issues; however, we hope that in short-term, they might help developers to survive in the industry.

\section{Threats}
\label{sec:threats}

As our focus was on understanding experiences and survival strategies of veteran software developers of marginalized genders, threats to \emph{construct validity} stem from our operationalization of such constructs as ``veteran'' and ``software developer''. 
We have used a cut-off of 18 years to identify veterans. However, our interview with Riley, who had 16 years of experience and whose experiences and survival strategies concurred with those of other interviewees, suggest that the exact value chosen as a threshold did not threaten the operationalisation of the ``veteran'' construct.

As observed in both this work and in the literature~\cite{Baltes20}, developers tend to change their jobs as they get older, e.g., moving to consultancy or management.  Hence, focusing solely on the individuals that, after an extended period of time, are still employed as software developers would have biased our study. This is why, in our study, we treat ``software development'' as an industry and include people who have moved into other roles within the industry. Furthermore, while we focus on individuals who are still active in software development, comparison of the interviews of Elliot (left) and Noah (returned) to the other interviewees suggests that the experiences and strategies identified have also been used by developers who (temporarily) left the industry after a career in software development. This means that the strategies may not be sufficient to retain marginalized genders in the industry. However, we do not argue that individuals adopting these strategies will solve systemic problems, and instead present them to highlight the extent to which remaining in the industry requires the adoption of techniques to overcome bias. 

An example of a threat to \emph{internal} validity is selection bias.
Interviewees have been selected among individuals that have participated in Twitter discussions.
Moreover, they have self-selected by agreeing to participate in the study.
However, some degree of self-selection is unavoidable in any interview study as prospective interviewees can always decline to participate.
Given the topic of the study, it might have been the case that prospective participants that did not have strong negative or positive gender- or age-related experiences decided not to participate, suggesting that the experiences reported are ``extreme'' in some sense.
However, many findings related to experience of \change{veteran women} developers reported in the current study concur with findings reported in the literature for women in software development, older people in software development, or gender/age related issues in the workforce in general. Future work could strengthen the findings by surveying software developers on the strategies they employ in order to identify to what extent they are widespread, and both age- and gender-dependent.

\emph{External} validity concerns generalization of the findings. In a qualitative study, we expect to establish theoretical generalization, which can be achieved through purposive sampling.
Having identified age as one of the key concerns of our work, we sought to interview people from different age groups within our target population. 
This demonstrates that our findings do not just apply to a subset of the population.
While we were able to include at least one participant from other demographic categories which could potentially affect the findings, future work should explicitly consider factors such as race and culture. 
The strategies we present include some not commonly seen in the literature, and all strategies were recognized by attendees, suggesting that these strategies are employed beyond our sample.
\change{Our sampling strategy based on Twitter and the fact that our participants all come from the Global North might bias our results towards privileged experiences. Future work needs to also consider the experience of veteran women developers beyond that group.}
\change{We presented partial results to a group of (mostly older women) practitioners in the UK. We also shared a preprint with two veteran women software developers working for a large international software company in Germany. 
Many of the reported phenomena were confirmed to resemble either their own experience or the experience of women they know.
These interactions allowed us to disseminate our results beyond veteran women developers working in North America, confirming that the phenomena we observed exists beyond that group.}
\change{Moreover, while we initially 
targeted all marginalized genders, we acknowledge that our findings primarily generalize to women, as the single non-binary participant identified as a woman for much of their career.} 

Finally, \emph{conclusion} validity relates to the ability to draw conclusions about the relationship between treatment and the results. It is possible that information was missed due to interpretation, as the coding was done by one author. However, the completed analysis and several code segments were reviewed by all authors, one of whom is a woman who left the software industry. Because of the complexity of the intersection of age and gender, the decision to code certain segments as pertaining to age, to gender, or to age and gender, was based on explicit references to these factors, and consequently the categorization may not have fully reflected the participants' experiences. As we noted in the introduction, this is a frequent problem when analyzing intersectional experiences. Nonetheless, we believe that the experiences, perspectives, and strategies portrayed are an authentic representation of the lived experiences of veteran \change{women} in the software industry. 
\section{Conclusion}
\label{sec:conclusion}
The intersection of ageism and sexism can create a hostile environment for veteran software developers of marginalized genders. 
This intersection \change{makes} experiences of older \change{women} software developers unique, and hence these experiences cannot be explained by merely \change{a} combination of experiences related to gender and age in isolation.
Thus, in this paper we have conducted a series of semi-structured interviews gauging the experiences and survival strategies of people working in the software industry whose gender and age are marginalized.
Our findings show that, indeed, gender- and age-related experiences often cannot be easily separated, and that our interviewees employed many different and diverse strategies, ranging from adjusting appearance to specialisation, and from speaking up to engaging in side projects.
Based on these findings, we call organizations in the software industry to understand the issues veteran women are facing and to act upon this understanding.
Organisations should invest in creating a good working environment and a positive atmosphere, investing \change{veteran women} developers with sense of control of their work and careers, supporting their promotion, assigning tasks and paying them on par with men.
\change{Researchers can use our results to derive hypotheses regarding the experiences related to age, gender, or their intersection. To differentiate those dimensions further, a comparative study involving veteran women, but also younger women and veteran non-women, is required.}

\section*{Acknowledgements}
The authors would like to thank the interviewees for participating in our study, the reviewers for their constructive feedback, and Eleni Constantinou and George Fletcher for their feedback on an early version of this manuscript.

\clearpage
\balance
\bibliographystyle{IEEEtran}
\bibliography{references}

\end{document}